\documentclass[aps, pra, 10pt, twocolumn, tightenlines, letterpaper, amsmath, amssymb, preprintnumbers, floatfix, longbibliography, nofootinbib]{revtex4-2}

\usepackage{dsfont}
\usepackage{amsmath}
\usepackage{amssymb}
\usepackage{physics}
\usepackage{tabu}
\usepackage{tabularx}
\usepackage{bm}
\usepackage{tikz}
\usetikzlibrary{shapes.geometric}
\usepackage{graphicx}
\usepackage{placeins}
\usepackage{textgreek}
\usepackage{multirow}
\usepackage{makecell}
\usepackage{soul}
\usepackage{diagbox}
\usepackage{xcolor}
\usepackage[normalem]{ulem}
\usepackage[export]{adjustbox}
\usepackage{float}
\usepackage{wasysym}

\newcommand{\bs}{\boldsymbol}
\newcommand{\nn}{\nonumber}

\newcommand{\cellHHC}{\scalebox{0.008}{\begin{tikzpicture}[baseline=(current bounding box.center)]
    \node (A) at (0,0) [] {};
    \node (B) at (3.7,-1.9) [] {};
    \node (C) at (18.5,-2.2) [] {};
    \node (D) at (25.6,10.8) [] {};
    \node (D) at (25.6,10.8) [] {};
    \node (E) at (17.9,24.2) [] {};
    \node (F) at (3.2,24.4) [] {};
    \node (G) at (-0.5,26.3) [] {};
    \node (H) at (10.6,28.0) [] {};
    \node (I) at (25.3,27.7) [] {};
    \node (J) at (32.9,14.4) [] {};
    \node (K) at (47.6,14.0) [] {};
    \node (L) at (51.4,12.1) [] {};
    \node (M) at (40.3,10.5) [] {};
    \node (N) at (25.8,1.4) [] {};
    \node (O) at (11.1,1.7) [] {};
    \draw (A) -- (B) -- (C) -- (D) -- (E) -- (F) -- (G) -- (H) -- (I) -- (J) -- (K) -- (L) -- (M) -- (D);
    \draw (J) -- (N) -- (O) -- (A);
\end{tikzpicture}}}

\newcommand{\loopHHCb}{\scalebox{0.008}{\begin{tikzpicture}[baseline=(current bounding box.center)]
    \node (D) at (25.6,10.8) [] {};
    \node (E) at (17.9,24.2) [] {};
    \node (F) at (3.2,24.4) [] {};
    \node (G) at (-0.5,26.3) [] {};
    \node (H) at (10.6,28.0) [] {};
    \node (I) at (25.3,27.7) [] {};
    \node (J) at (32.9,14.4) [] {};
    \node (K) at (47.6,14.0) [] {};
    \node (L) at (51.4,12.1) [] {};
    \node (M) at (40.3,10.5) [] {};
    \draw (D) -- (E) -- (F) -- (G) -- (H) -- (I) -- (J) -- (K) -- (L) -- (M) -- (D);
\end{tikzpicture}}}

\usepackage[hypertexnames=false]{hyperref}
\hypersetup{
    colorlinks=true,       
    linkcolor=blue,          
    citecolor=blue,        
    filecolor=blue,      
    urlcolor=blue           
}

\newcolumntype{Y}{>{\centering\arraybackslash}X}

\usepackage{orcidlink}

\begin{document}

\begin{figure}
  \vskip -1.cm
  \leftline{\includegraphics[width=0.15\textwidth]{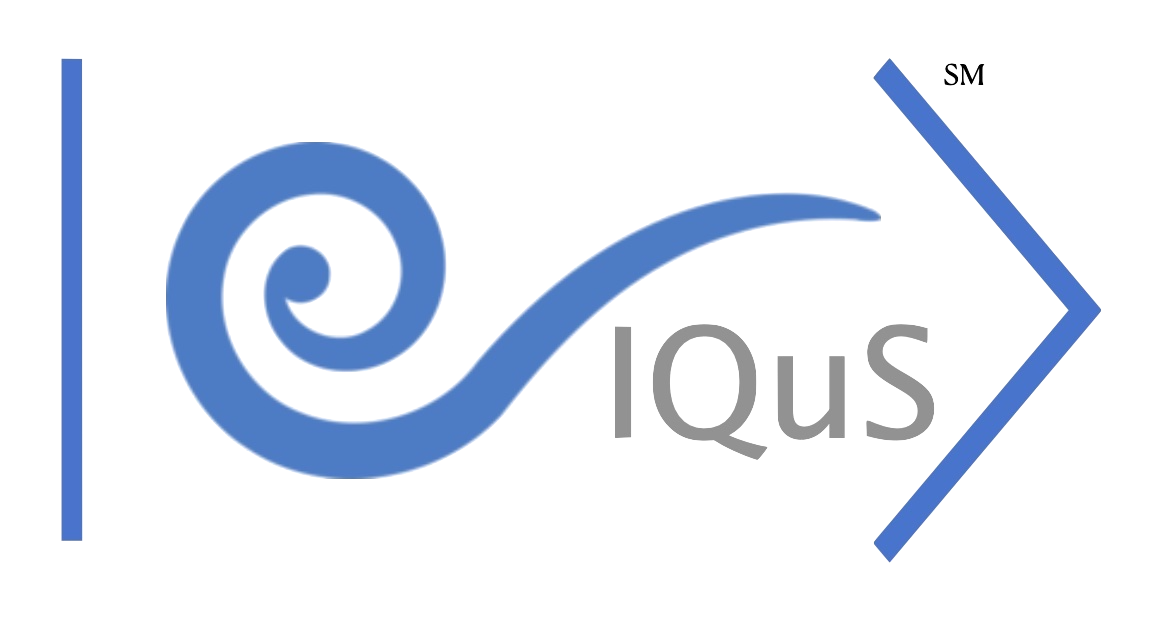}}
\end{figure}

\title{
Improved Honeycomb and Hyperhoneycomb Lattice Hamiltonians for Quantum Simulations of Non-Abelian Gauge Theories}

\author{Marc Illa\,\orcidlink{0000-0003-3570-2849}}
\email{marcilla@uw.edu}
\affiliation{InQubator for Quantum Simulation (IQuS), Department of Physics, University of Washington, Seattle, WA 98195, USA.}

\author{Martin J.~Savage\,\orcidlink{0000-0001-6502-7106}}
\email{mjs5@uw.edu}
\thanks{On leave from the Institute for Nuclear Theory.}
\affiliation{InQubator for Quantum Simulation (IQuS), Department of Physics, University of Washington, Seattle, WA 98195, USA.}

\author{Xiaojun Yao\,\orcidlink{0000-0002-8377-2203}}
\email{xjyao@uw.edu}
\affiliation{InQubator for Quantum Simulation (IQuS), Department of Physics, University of Washington, Seattle, WA 98195, USA.}

\preprint{IQuS@UW-21-097}
\date{\today}

\begin{abstract}
\noindent
Improved Kogut-Susskind Hamiltonians for quantum simulations of  non-Abelian Yang-Mills gauge theories are developed for honeycomb (2+1D) and hyperhoneycomb (3+1D) spatial tessellations.
This is motivated by the desire to identify lattices for quantum simulations that involve only 3-link vertices among the gauge field group spaces in order to reduce the complexity in applications of the plaquette operator. 
For the honeycomb lattice, we derive a classically ${\cal O}(b^2)$-improved Hamiltonian, with $b$ being the lattice spacing.  
Tadpole improvement via the mean-field value of the plaquette operator is used to provide the corresponding quantum improvements.
We have identified the (non-chiral) hyperhoneycomb as a candidate spatial tessellation for 3+1D quantum simulations of gauge theories, and determined the associated ${\cal O}(b)$-improved Hamiltonian.
\end{abstract}

\maketitle

\section{Introduction}
\label{sec:intro}
\noindent
Quantum simulations of non-Abelian quantum field theories are a key ingredient in establishing robust predictions for the real-time dynamics in non-equilibrium processes important to nuclear physics and high-energy physics (for recent reviews, see Refs.~\cite{Banuls:2019bmf,Klco:2021lap,Bauer:2022hpo,Beck:2023xhh,DiMeglio:2023nsa,Bauer:2023qgm}). 
They will provide access to aspects of fundamental physics that lie far beyond the capabilities of analytic calculation and classical computing alone~\cite{5392446,5391327,doi:10.1063/1.881299,Manin1980,Benioff1980,
Feynman1982,Feynman1986,Preskill2018quantumcomputingin}, providing, for example, unique insights into the evolution of matter under extreme conditions in the early universe and supernova, 
and foundational improvements to rates and cross sections of low-energy nuclear reactions involving short-lived isotopes relevant for nuclear energy generation.
While there is no unique path forward for simulating such processes using near-term quantum computers, the Kogut-Susskind (KS) Hamiltonian formulation~\cite{Kogut:1974ag,Banks:1975gq} of Yang-Mills lattice gauge theories (LGT) is one of the most promising formulations. 
Early quantum simulations of non-Abelian LGTs
beyond 1+1D have been performed using the KS Hamiltonian with NISQ-era quantum computers, using the more traditional cubic tessellation of spacetime~\cite{Klco:2019evd, Ciavarella:2021nmj,Ciavarella:2021lel,ARahman:2021ktn,Illa:2022jqb,ARahman:2022tkr,Ciavarella:2024fzw}.
Further, significant formal developments toward optimizing mappings and constraint implementations are in process,
e.g., Refs.~\cite{Bronzan:1984xb,Bronzan:1991pa,Davoudi:2020yln,Bauer:2021gek,Hartung:2022hoz,Grabowska:2022uos,DAndrea:2023qnr,Alexandru:2023qzd,Gustafson:2023kvd,Gustafson:2024kym,Fontana:2024rux,Grabowska:2024emw,Halimeh:2024bth}.
One of the well-known challenges in simulating non-Abelian field theories, in contrast to Abelian theories, is the inclusion of the Lie-algebra group-space 
wavefunction.
While this mapping problem was addressed in the pioneering work of Byrnes and Yamamoto~\cite{Byrnes_2006}, and subsequent suggested implementations, e.g., Refs.~\cite{Banuls:2017ena,Alexandru:2019nsa,Ciavarella:2021nmj}, a significant complication is in the coupling of link group spaces at each of the lattice sites.
While in 1+1D this problem is straightforwardly dealt with, uniquely constrained by Gauss's law, in 2+1D this requires recoupling the group spaces of four links of a square lattice for each application of the plaquette operator, and six links in 3+1D.
In the electric basis, this becomes increasingly challenging and a limitation with decreasing coupling as higher group representations provide support to the wavefunction.
This has led to the development of alternate mappings, all of which share the same continuum limit, 
including the loop-string-hadron framework~\cite{Raychowdhury:2018osk,Raychowdhury:2019iki,Kadam:2022ipf,Kadam:2024zkj}, 
controlled plaquette operators~\cite{Ciavarella:2021nmj},
using qudit quantum computers~\cite{Gonzalez-Cuadra:2022hxt,Gustafson:2023swx,Zache:2023cfj,Meth:2023wzd,Calajo:2024qrc,Illa:2024kmf,Araz:2024kkg}, 
q-deformed truncations in gauge space~\cite{Zache:2023dko},
or directly mapping to tri-coordinate lattices, such as honeycomb lattices in 2+1D~\cite{Muller:2023nnk,Turro:2024pxu,Turro:2025sec},
and triamond lattices in 3+1D~\cite{Kavaki:2024ijd,Kavaki:2025hcu}.\footnote{
The triamond lattice can be denoted as a (10,3)a network using the Schl\"{a}fli symbols, a K4 crystal, or hyperoctagon lattice. For a review, see Ref.~\cite{Sunada2008}.
}\textsuperscript{,}\footnote{
4D hyperdiamond lattices have been previously considered in the context of Euclidean-space lattice QCD simulations~\cite{Bedaque:2008jm,Kimura:2009qe,Kimura:2009di}, in an
effort to formulate chiral fermions without fine-tunings.
}
Many of these implementations establish simulation protocols that involve (re)coupling only three link spaces in any given application of the plaquette operator, 
rendering this aspect of the simulation (more) manageable.

Pursuing this line of development for quantum simulating non-Abelian LGTs,
in this work, 
we establish lattice-spacing improved Hamiltonians 
for the honeycomb (HC) tessellation of 2+1D spacetime and  
the hyperhoneycomb (HHC) tessellation\footnote{
The HC is also denoted as (6,3), and HHC as (10,3)b.
}
of 3+1D spacetime following the Symanzik improvement program~\cite{Symanzik:1983dc}, similar to 
previous results for cubic tessellation in Refs.~\cite{Moore:1996wn,Luo:1998dx,Carlsson:2001wp,Carlsson:2003rf}.\footnote{
There have been recent advances focused on implementing improvements for cubic lattices using quantum computers, e.g., Refs.~\cite{Carena:2022kpg,Ciavarella:2023mfc,Gustafson:2023aai}.
}
Part of the motivation for this program is to further connections between the 
lattice simulations of non-Abelian gauge theories for fundamental physics, 
which necessarily involve truncations in the group space, 
and quantum spin liquids (QSL), which play an important role in quantum information and computing (for a review, see Ref.~\cite{Savary_2016}), 
exemplified by Kitaev's exactly solvable HC model~\cite{Kitaev:2005hzj} 
and extensions into 3D~\cite{Si:2007,Si:2008,Mandal:2009,OBrien:2016,Jahromi_2021} (for reviews, see, e.g., Refs.~\cite{doi:10.7566/JPSJ.89.012002,TREBST20221}).
QSL phases typically have short-range correlations, long-range entanglement,
fractionalized excitations, and topological order.
For a range of couplings and group-space truncations, non-Abelian LGTs exhibit a QSL phase
which diminishes rapidly with increasing truncation~\cite{Zache:2023dko}.  
In addition to the advantages from 3-link vertices, establishing further parallels between 
non-Abelian LGTs of the Standard Model and QSL could be beneficial.

\section{Definitions}
\label{sec:definitions}
\noindent
The   Hamiltonian for pure SU$(N_c)$ Yang-Mills in the continuum can be written as
\begin{align}
\label{eqn:H_YM}
       H_{\rm YM} & = \int d{\bs x} \bigg\{ \frac{1}{2} \sum_{a,i}[E^a_i({\bs x})]^2 + \frac{1}{2} \sum_{a,i} [B^a_i({\bs x})]^2 \bigg\}
       \nn\\
       & = H_E + H_B \ , 
\end{align}
with  $E^a_i$ and $B^a_i$ being the chromo-electric and chromo-magnetic fields, respectively, in the $i^{\rm th}$ spatial direction, and $a$ being the SU$(N_c)$ adjoint index running over $[1,\ldots,N^2_c-1]$. 
In terms of the field-strength tensor $G^a_{\mu\nu} = \partial_\mu A^a_\nu - \partial_\nu A^a_\mu - g f^{abc}A^b_\mu A^c_\nu$, where $g$ is the coupling constant, $A^a_\mu$ is the vector potential, and $f^{abc}$ is the structure constant, the chromo-electric and chromo-magnetic fields are $E^a_i=G^a_{0i}$ and $B^a_i=-\frac{1}{2}\epsilon_{ijk}G^a_{jk}$. Covariant derivatives acting on the field tensor are defined as $D_\mu G_{\mu\nu}=\partial_\mu G_{\mu\nu} +ig[A_{\mu},G_{\mu\nu}]$.

\subsection{Gauge fixing}
\label{subsec:gaugefix}
\noindent
Wilson loops~\cite{Wilson:1974sk} are
one of the building blocks of discretized Hamiltonians, also known as plaquettes, which are products of link variables $U$ (Wilson lines) along a closed loop $\mathcal{C}$,
\begin{equation}
    P = \prod_{k\in \mathcal{C}}U({\bs x}_k, {\bs e}_k) \ .
    \label{eq:plaq}
\end{equation}
A general link variable is defined as,
\begin{equation}
U({\bs x}, {\bs e}) = \mathcal{P}\exp\left[ ig \int_{\bs x}^{{\bs x}+{\bs e}} dz_i A_i({\bs z})\right] 
\ ,
\end{equation}
where  $\mathcal{P}$ denotes path ordering and the summation is over spatial indices $i$. 
The path of the Wilson line is a straight line starting at position ${\bs x}$, pointing along ${\bs e}$ and ending at ${\bs x}+{\bs e}$.

The plaquette operator in Eq.~\eqref{eq:plaq} is defined in a gauge-invariant way. 
However, if one naively expands it with respect to a spatial point in powers of the lattice spacing $b$, 
the results are not necessarily gauge invariant order-by-order in $b$. 
This is because, for example, in 2+1D, 
there is only one physical gauge degree of freedom. 
The plaquette operator depends on both $A_x$ and $A_y$, and naively expanding Eq.~\eqref{eq:plaq} in powers of $b$ treats both $A_x$ and $A_y$ as independent degrees of freedom. As a result, different equivalent gauge orbits mix with each other. 
For example, at ${\cal O}(b^5)$, a ${\cal O}(b^2)$ term from one gauge orbit multiplying 
a ${\cal O}(b^3)$ term from another gauge orbit can contribute, which spoils gauge invariance at order ${\cal O}(b^5)$.
This problem in the expansion, caused by the gauge redundancy, is cured by gauge fixing. 
One gauge-fixing choice that significantly simplifies calculations was used in Ref.~\cite{Luscher:1984xn}. 
Modifying this gauge choice for the 2+1D KS Hamiltonian setup, 
where $A_0=0$ has been used everywhere in spacetime, 
we use the following additional spatial gauge-fixing conditions,
\begin{align}
A_y(x,y) &= 0 \ \ \forall \ \ x,y \ , \nn \\
A_x(x,y) &= 0 \ \ \forall \ \ x\ \  {\rm at}\ \ y=0 \ .
\end{align}
As a result of this gauge-fixing choice, 
\begin{align}
\partial_x^n A_x(0,0) &= 0 \ \ \forall \ \ n \in \mathbb{Z}_{\geq 0} \ , \nn\\
\partial_x^n \partial_y^{m+1}  A_x(0,0) &= D_x^n D_y^{m} G_{yx}(0,0) \ \ \forall \ \ n,m \in \mathbb{Z}_{\geq 0} \ , \nn\\
\partial_x^n \partial_y^{m}  A_y(0,0) &= 0 \ \ \forall \ \ n,m \in \mathbb{Z}_{\geq 0} \ . 
\end{align}

In 3+1D, we use the following specific gauge fixing,
\begin{align}
A_z(x,y,z) & = 0 \ \ \forall \ \ x,y,z\ , \nn\\
A_x(x,y,z) & = 0 \ \ \forall \ \ x,y\ \  {\rm at}\ \ z=0 \ , \nn\\
A_y(x,y,z) & = 0 \ \ \forall \ \ y\ \  {\rm at}\ \ x=z=0 \ ,
\label{eq:HHCgf}
\end{align}
which results in
\begin{align}
\partial_x^n\partial_y^m A_x(0,0,0) &= 0 \ \ \forall \ \ n,m \in \mathbb{Z}_{\geq 0} \ , \nn\\
\partial_x^n\partial_y^m\partial_z^{l+1} A_x(0,0,0) &= D_x^n D_y^{m}D_z^{l} G_{zx} \ \ \forall \ \ n,m,l \in \mathbb{Z}_{\geq 0} \ , \nn\\
\partial_y^{m}  A_y(0,0,0) &= 0 \ \ \forall \ \ m \in \mathbb{Z}_{\geq 0} \ , \nn \\
\partial_x^{n+1}\partial_y^m\partial_z^l A_y(0,0,0) &= D_x^n D_y^{m}D_z^{l} G_{xy} \ \ \forall \ \ n,m,l \in \mathbb{Z}_{\geq 0} \ , \nn\\
\partial_x^{n}\partial_y^m\partial_z^{l+1} A_y(0,0,0) &= D_x^n D_y^{m}D_z^{l} G_{zy} \ \ \forall \ \ n,m,l \in \mathbb{Z}_{\geq 0} \ , \nn\\
\partial_x^{n}\partial_y^m\partial_z^l A_z(0,0,0) &= 0 \ \ \forall \ \ n,m,l \in \mathbb{Z}_{\geq 0} \ .
\end{align}
From the second and third to last lines, 
in this specific gauge condition, we find
\begin{equation}
D_x^nD_y^m D_z^{l+1} G_{xy} = D_x^{n+1}D_y^m D_z^{l} G_{zy} \ .
\label{eq:gaugegzy}
\end{equation}

\subsection{Path ordering}
\label{subsec:path}
\noindent
To obtain a systematic expansion, 
the line integral in the exponent is expanded in powers of displacement from  a chosen vertex. 
Furthermore, due to the path ordering in the definition of the link operator, which is crucial for the gauge covariance, 
the exponential function is expanded in powers of displacement, 
then path-ordering is applied, 
followed by integration over the displacement between adjacent vertices.\footnote{
To illustrate this point, consider a simple link operator in 1D, 
given by
\begin{equation}
\ell_1 = \mathcal{P}\exp\left[ ig\int_0^b dx A(x)  \right] \ .
\end{equation}
Expanding the integral in the exponent first gives, at ${\cal O}(b^3)$,
\begin{equation}
\ell_1 \rightarrow b^3\left[ \frac{i}{6} \partial_x^2 A -\frac{1}{4} A\partial_x A - \frac{1}{4}(\partial_x A)A -\frac{i}{6} A^3 \right] \ ,
\end{equation}
while expanding the exponential function first gives, at ${\cal O}(b^3)$,
\begin{equation}
\ell_1 \rightarrow b^3\left[ \frac{i}{6} \partial_x^2 A -\frac{1}{3} A\partial_x A - \frac{1}{6}(\partial_x A)A -\frac{i}{6} A^3 \right]\,,
\end{equation}
where the path ordering is accounted for. 
The coefficients of $A\partial_x A$ and $(\partial_x A)A$ are different, 
demonstrating the importance of path ordering in expanding link operators.
}

\section{Honeycomb Lattices for 2+1D Simulations}
\label{sec:HC}
\noindent
Mapping the KS Hamiltonian onto a honeycomb lattice was first implemented in Ref.~\cite{Muller:2023nnk}, as mentioned in the Introduction. 
In this section we will first review and recover the leading-order results, and extend those to suppress lattice spacing artifacts.

\begin{figure}
    \centering
    \includegraphics[width=0.75\linewidth]{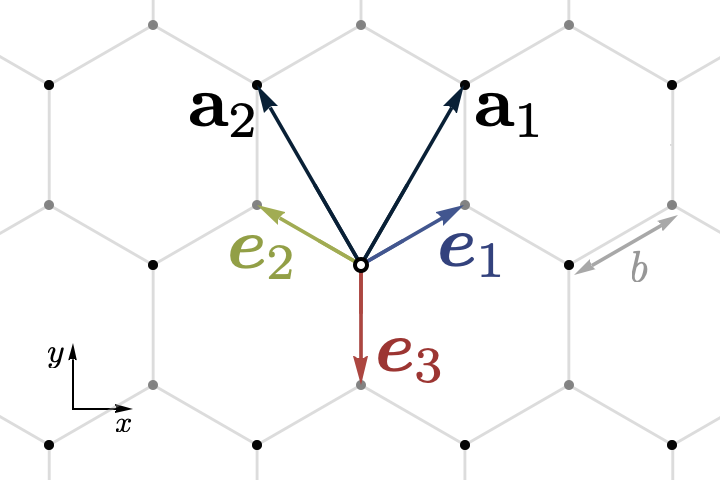}
    \caption{The honeycomb lattice defined by the vectors in Eq.~\eqref{eq:HClinks}, with the colored links highlighting a unit cell.}
    \label{fig:HCbaselatt}
\end{figure}

The HC lattice (see Fig.~\ref{fig:HCbaselatt}) can be defined by three link directions ${\bs e}_i$ (unit cell) and two lattice vectors ${\bf a}_i$ (translation vectors), in units of the length of each link $b$,
\begin{align}
    {\bs e}_1 & = \frac{1}{2}(\sqrt{3}, 1 ) \ , \ {\bs e}_2 = \frac{1}{2}( -\sqrt{3}, 1 ) \ , \ {\bs e}_3 = (0,-1) \ , \nn \\
    {\bf a}_1 & = \frac{1}{2}(\sqrt{3}, 3 ) \ , \ {\bf a}_2 = \frac{1}{2}(-\sqrt{3}, 3 ) \ ,
    \label{eq:HClinks}
\end{align}
with any point on the lattice being generated by shifting the unit cell along the lattice vectors, ${\bs x}= n {\bf a}_1 + m {\bf a}_2$, 
with $n,m\in \mathbb{Z}$. The center of the lattice is the white point in Fig.~\ref{fig:HCbaselatt}, and it will be the origin of all the subsequent expansions. A relevant quantity is the area of a single cell, 
\begin{equation}
  S_{\varhexagon} = |{\bf a}_1\times {\bf a}_2| = 3\sqrt{3}b^2/2 \ .
  \label{eq:hexarea}
\end{equation}

\subsection{Magnetic contribution}
\label{subsec:HCmag}
\noindent
The magnetic contribution to the Hamiltonian is computed from closed loops running around the links of the HC lattice. 
For square lattices, square and rectangle loops are combined to systematically reduce the lattice spacing artifacts~\cite{Luscher:1984xn}.

\begin{figure}
    \centering
    \includegraphics[width=0.95\linewidth, valign=c]{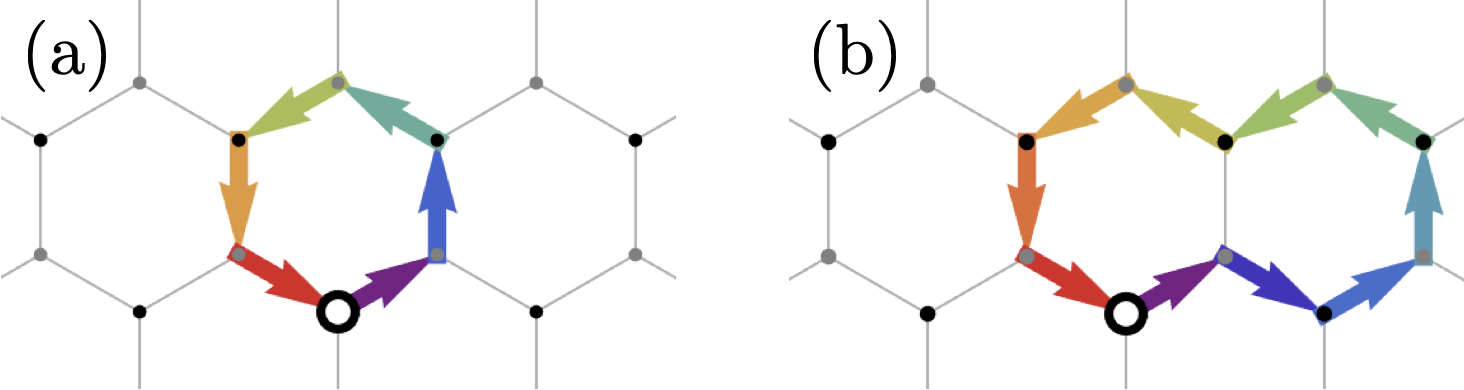}
    \caption{Examples of (a) 6-link and (b) 10-link loops on the HC lattice.}
    \label{fig:HC610}
\end{figure}
On the HC lattice, the smallest plaquette loop operator is given by a 6-link loop (more details can be found in Appendix \ref{app:HC}).  
An example is shown in Fig.~\ref{fig:HC610}(a), and is computed via
\begin{equation}
\hat{P}_{6}({\bs x}) = {\rm Tr}\left[  U_1 U_2 U_3 U_4 U_5 U_6 \right] \ ,
\label{eq:hc_plaq6}
\end{equation}
where the link operators are defined by
\begin{align}
U_1 &= U({\bs x}, {\bs e}_1) \ , & \quad U_2 &= U^\dagger({\bs x}+{\bf a}_1, {\bs e}_3) \ , \nn\\
U_3 &= U({\bs x}+{\bf a}_1, {\bs e}_2) \ , & \quad U_4 &= U^\dagger({\bs x}+{\bf a}_2, {\bs e}_1) \ , \nn \\
U_5 &= U({\bs x}+{\bf a}_2, {\bs e}_3) \ , & \quad U_6 &= U^\dagger({\bs x}, {\bs e}_2) \ .
\end{align}
Using the gauge-fixing and expansions detailed in Sec.~\ref{sec:definitions}, each loop can be written as $\hat{P}_{6}({\bs x}) = {\rm Tr}[e^{ig\Phi(\bs x)}]$, with $\Phi(\bs x)$ having the following general form,\footnote{
One may wonder why the expansion of the HC plaquette can be organized in terms of $G_{xy}$, $D_x$ and $D_y$, which are natural on the square lattice and obey certain rotation rules under the square rotation group $C_4$. 
On the HC lattice, the natural gauge-field components are 
$A_\alpha={\bs e}_\alpha \cdot {\bs A}$, with 
${\bs A}=(A_x,A_y)$ and $\alpha\in[1,2,3]$. The natural field-strength tensors $G^a_{\alpha\beta}$ are defined in terms of $A_\alpha$ and 
$\partial_\alpha={\bs e}_\alpha \cdot 
{\bs\partial}$, e.g., $G^a_{12}=\partial_1 A_2 - \partial_2 A_1 - gf^{abc}A_1^bA_2^c$. 
We find $G^a_{12}=G^a_{23}=G^a_{31}=\frac{\sqrt{3}}{2}G^a_{xy}$. 
At $\mathcal{O}(b^2)$ in the expansion of $\hat{P}_6$, $(G^a_{12})^2$, $(G^a_{23})^2$ and $(G^a_{31})^2$ are all equivalent to $(G_{xy}^a)^2$. 
At $\mathcal{O}(b^3)$, we find the three loops in Fig.~\ref{fig:HC6} are $\propto D_\alpha G_{xy}^a$, where $D_\alpha={\bs e}_\alpha \cdot {\bs D}$. 
These expressions are also natural to the HC lattice, 
since $G^a_{\alpha\beta}=\pm\frac{\sqrt{3}}{2}G^a_{xy}$.}
\begin{align}
    \Phi(\bs x) = &\  S_{\varhexagon} \left[\alpha G_{xy}(\bs x)+b(\beta_1 D_x + \beta_2 D_y)G_{xy}(\bs x) \right. \nn \\
    & \left. + b^2(\gamma_1 D^2_x + \gamma_2 D_xD_y +\gamma_3 D^2_y)G_{xy}(\bs x) +\mathcal{O}(b^3) \right] \ ,
    \label{eq:HCloop}
\end{align}
with the numerical value of the coefficients $\{\alpha, \beta_i, \gamma_i\}$ for each loop given in Table~\ref{tab:HCcoeffs} in Appendix \ref{app:HC}. 
After expanding and averaging over these 6-link  loops,
\begin{align}
    & \Gamma^{({\rm HC})}_6 = \frac{1}{N_{\Phi}}\sum_i{\rm Tr}[\hat{I}-e^{ig\Phi_i}] \label{eq:P6exp}\\
    &= \frac{g^2}{2}S^2_{\varhexagon}{\rm Tr}\left[G^2_{xy}+\frac{5}{24}b^2 G_{xy}(D_x^2+D_y^2)G_{xy}+\mathcal{O}(b^3) 
    \right] \ , \nn
\end{align}
with $N_{\Phi}=3$ being the number of loops.
As in the case with ``clover'' averaging in the cubic lattice~\cite{GarciaPerez:1993lic}, the contribution linear in the covariant derivative vanishes.

The contribution quadratic in the covariant derivative can be removed by including 10-link loops, which are the next largest after the 6-link loops, see Fig.~\ref{fig:HC610}(b). 
While there are 15 different loops of this kind, 9 of them are sufficient (see Appendix \ref{app:HC}). 
These can be expanded in a similar form of Eq.~\eqref{eq:HCloop}, 
leading to (after averaging),
\begin{align}
    \Gamma^{({\rm HC})}_{10} = \frac{g^2}{2} S^2_{\varhexagon} {\rm Tr}\bigg[ 4 G^2_{xy} +\frac{7}{3}b^2G_{xy}(D^2_x +D^2_y)G_{xy} \nn\\
    + \mathcal{O}(b^3)\bigg] \ .
\end{align}
Combining both types of loops with appropriate coefficients leads to, 
\begin{align}
    \Gamma^{({\rm HC})}_{6,10} &= \frac{14}{9}\Gamma^{({\rm HC})}_{6} -\frac{5}{36}\Gamma^{({\rm HC})}_{10} \nn \\
    & = \frac{g^2}{2}S^2_{\varhexagon}{\rm Tr}[G^2_{xy}]+\mathcal{O}(b^7) \nn \\
    & = \frac{g^2}{4}S^2_{\varhexagon}\sum_a [B^a({\bs x})]^2+\mathcal{O}(b^7) \ .
    \label{eq:Bhcimp}
\end{align}
Then, the $\mathcal{O}(b^2)$ improved magnetic contribution to the Hamiltonian is given by
\begin{equation}
    H_B = 
    \frac{S_{\varhexagon}}{2}
    \sum_{a,{\bs x}}[B^a({\bs x})]^2 \ 
    \rightarrow \ 
    \frac{2}{g^2S_{\varhexagon}}
    \sum_{\rm tiles}
    \Gamma^{({\rm HC})}_{6,10} 
    \ .
    \label{eq:Bhcimp2}
\end{equation}

\subsection{Electric contribution}
\label{subsec:HCe}
\noindent
As noted in Refs.~\cite{Moore:1996wn,Luo:1998dx,Carlsson:2001wp,Carlsson:2003rf}, the electric term in the Hamiltonian has to be improved as well.
In particular, from Refs.~\cite{Carlsson:2001wp,Carlsson:2003rf}, the general form for the $\mathcal{O}(b^2)$ improved electric term in a $d$+1D cubic lattice is given by
\begin{align}
& \int d{\bs x}\, \frac{1}{2} \sum_{a,i} [E^a_i({\bs x})]^2 = \int d{\bs x}\, \sum_{i} {\rm Tr}\left[ | E_i({\bs x})|^2 \right] \nn\\
& \  \rightarrow  b^d \sum_{i,{\bs x}} 
{\rm Tr}\left[ \left(1-\frac{1}{6n^2}\right)
|E_i({\bs x})|^2 
\right.
\label{eq:Elimp}
\\
& \left. \qquad  + \ 
\frac{1}{6n^2} 
 E_i({\bs x}) U({\bs x},n{\bf a}_i) E_i({\bs x}+n{\bf a}_i) U^\dag({\bs x},n{\bf a}_i)
\right] \ , \nn
\end{align}
with $n$ being the number of links away from ${\bs x}$ that are used to smear the local electric field.

The electric field, ${\bs E}=(E_x, E_y)$, is mapped to the HC lattice~\cite{Muller:2023nnk}  via
\begin{align}
E_\alpha & = {\bs e}_\alpha\cdot {\bs E}  \ ,
\nn \\
E_1 & = \frac{1}{2} (\sqrt{3} E_x +  E_y) 
\ ,\ E_2 = \frac{1}{2} ( -\sqrt{3} E_x +  E_y) \ , \nn\\
E_3 & = -E_y \ ,
\label{eq:EHC}
\end{align}
where $E_\alpha$ denotes the value of the electric field along the link directions 
${\bs e}_\alpha$ with $\alpha=1,2,3$.
The magnitude of the electric field is given by
\begin{equation}
|{\bs E}|^2 = E_x^2+E_y^2 = \frac{2}{3} \sum_{\alpha=1}^3 E_\alpha^2 \ .
\label{eq:EHCcart}
\end{equation}
Therefore, the leading-order term in the electric Hamiltonian on the HC lattice is given by,
\begin{equation}
H_E = S_{\varhexagon} \sum_{i, {\bs x}} {\rm Tr}\left[|E_i({\bs x})|^2\right] \rightarrow \frac{2 S_{\varhexagon}}{3} \sum_{\substack{\alpha, {\rm cells}}} {\rm Tr}\left[|E_\alpha({\bs x})|^2\right] \ ,
\label{eq:HCloE}
\end{equation}
where the sum is over the HC links in each cell (black vertices in Fig.~\ref{fig:HCbaselatt}).
The classically-improved electric Hamiltonian
is of the form of Eq.~\eqref{eq:Elimp} with $n=3$,
\begin{align}
{\rm Tr}\left[|E_\alpha({\bs x})|^2\right] \rightarrow \ & \frac{53}{54} {\rm Tr}\left[|E_\alpha({\bs x})|^2\right]
\label{eq:elimphc} \\
& +  
\frac{1}{54} {\rm Tr} \left[
  E_\alpha({\bs x}) U_\alpha
  E_\alpha({\bs x}+3{\bs e_\alpha}) U_\alpha^\dagger \right]
\ , \nn
\end{align}
where $U_\alpha$ is a composite link operator connecting four vertices. 
For each $\alpha$, there are two possible 4-link paths, and while the particular path chosen will generate lattice spacing effects at even higher orders, possible choices for each lattice direction are 
\begin{align}
    U_1 = \ & U({\bs x},{\bs e}_1) U^\dag({\bs x}+{\bf a}_1,{\bs e}_3) \nn \\ & \times U({\bs x}+{\bf a}_1,{\bs e}_1) U^\dag({\bs x}+3{\bs e}_1,{\bs e}_2) \ , \nn \\
    U_2 = \ & U({\bs x},{\bs e}_2) U^\dag({\bs x}+{\bf a}_2,{\bs e}_3) \nn \\ & \times U({\bs x}+{\bf a}_2,{\bs e}_2) U^\dag({\bs x}+3{\bs e}_2,{\bs e}_1) \ , \nn \\
    U_3 = \ & U({\bs x},{\bs e}_3) U^\dag({\bs x}-{\bf a}_1,{\bs e}_1) \nn \\ & \times U({\bs x}-{\bf a}_1,{\bs e}_3) U^\dag({\bs x}+3{\bs e}_3,{\bs e}_2) \ .
\end{align}
Further parametric reductions in lattice-spacing artifacts will require a refined treatment along the lines of the analysis presented in Refs.~\cite{Carlsson:2001wp,Carlsson:2003rf} for square and cubic lattices.

\subsection{Tadpole improvement}
\label{subsec:HCTI}
\noindent
In 4D Euclidean space lattice gauge field theory simulations, it is well known that renormalizing the plaquette operator by the mean-field value of the links is essential for recovering accurate results~\cite{Lepage:1992xa}.   
This is because quantum fluctuations of the gauge field, particularly closed loops from the same 
point on the link (tadpole diagrams) give rise to UV divergences that cancel lattice spacing suppression factors to render a leading-order effect.
The norm of a link deviates from unity due to these quantum fluctuations, suppressed by powers of the strong coupling constant.  
In the Hamiltonian formulation, these effects have been considered previously and the appropriate renormalizations for a cubic lattice have been determined~\cite{Moore:1996wn,Luo:1998dx,Carlsson:2001wp,Carlsson:2003rf}.
In the context of quantum simulations, where dynamical processes are the focus, these mean-field renormalizations will typically be spacetime dependent, meaning local renormalizations will be required to be computed on each time slice.
The average renormalization of a single link,
$u_{\varhexagon}$, is estimated by
\begin{equation}
u_{\varhexagon}^6 = 1 + \frac{1}{2 N_c}\ \langle \psi |  \hat{P}_6 + \hat{P}_6^\dagger | \psi \rangle \ ,
\label{eq:hextpfac}
\end{equation}
where the 6-link plaquette operator is defined in
Eq.~\eqref{eq:hc_plaq6}.

\subsection{Improved Hamiltonian}
\label{subsec:HCIH}
\noindent
Hamiltonian improvement is a well-established technique for simulations to accelerate the approach to the continuum limit.   
Higher-order classical lattice spacing artifacts are removed in the weak-coupling limit by including contributions from larger loops to the magnetic Hamiltonian, and
gauge-invariant non-local smearing of the electric field operator.
In addition, as discussed in the previous section, quantum fluctuations that furnish leading-order effects are removed by ``tadpole-improvement'', facilitated by a state-dependent operator renormalization determined iteratively from a mean-field estimate of the link operator.
Using Eqs.~\eqref{eq:HCloE} and \eqref{eq:elimphc}, and rescaling the electric field $E_\alpha \to gE_\alpha/b$ as in the standard KS Hamiltonian, the electric contribution to the Hamiltonian is improved to
\begin{align}
H_E =
\frac{2g^2S_{\varhexagon}}{3b^2} & 
\sum_{\substack{\alpha, {\rm cells}}} 
 {\rm Tr}\Bigg[ \frac{53}{54}  |E_{\alpha} ({\bs x})|^2  \nn\\
& + \frac{1}{54 u_{\varhexagon}^8} 
  E_\alpha({\bs x}) U_\alpha
  E_\alpha({\bs x}+3{\bs e_\alpha}) U_\alpha^\dagger  
 \Bigg] \ ,
\end{align}
where there are four links contributing to $U_\alpha$, and as such the tadpole improvement factor for the electric field is $u_{\varhexagon}^4$ for each $U_\alpha$.
Improvement to the magnetic Hamiltonian involves renormalizing the action of the plaquette operator(s) with their mean-field values, 
together with Eqs.~\eqref{eq:Bhcimp} and \eqref{eq:Bhcimp2}, 
\begin{align}
H_B = 
\frac{1}{g^2S_{\varhexagon}}
\sum_{\rm tiles} &
\left[ \frac{17}{6} N_c -
\frac{14}{9 u_{\varhexagon}^6} \left(\hat P_{6} + \hat P_{6}^\dagger \right)
\right. \nn\\
& \ 
\left.
+ \, \frac{5}{36 u_{\varhexagon}^{10}} \left( \hat P_{10} + \hat P_{10}^\dagger \right)
\right]
\ .
\label{eq:classHCMagP}
\end{align}
%

\section{Hyperhoneycomb Lattices for 3+1D Simulations}
\label{sec:HHC}
\noindent
The hyperhoneycomb lattice is the 3+1D extension of the honeycomb lattice in 2+1D (and related via the harmonic honeycomb series~\cite{Modic:2014}) described in Sec.~\ref{sec:HC}.
Recent ground-breaking work by Kavaki and Lewis~\cite{Kavaki:2024ijd} suggested that triamond lattices could be a better 3+1D tessellation because of their 3-link vertices.   
We have examined these lattices at higher orders in the lattice spacing and find that, because of their intrinsic chirality and the additional operator structure that this engenders,\footnote{For example, the operator $\epsilon_{ijk}G_{il}D_jG_{kl}$.} 
the loop structure necessary for classically improvement is complicated. 
However, HHC tessellations are not chiral, leading to simpler higher-order structures,
and combinations of loops provide an improved Hamiltonian for 
3+1D simulations.

\begin{figure}[tb!]
    \centering
    \includegraphics[width=0.8\linewidth]{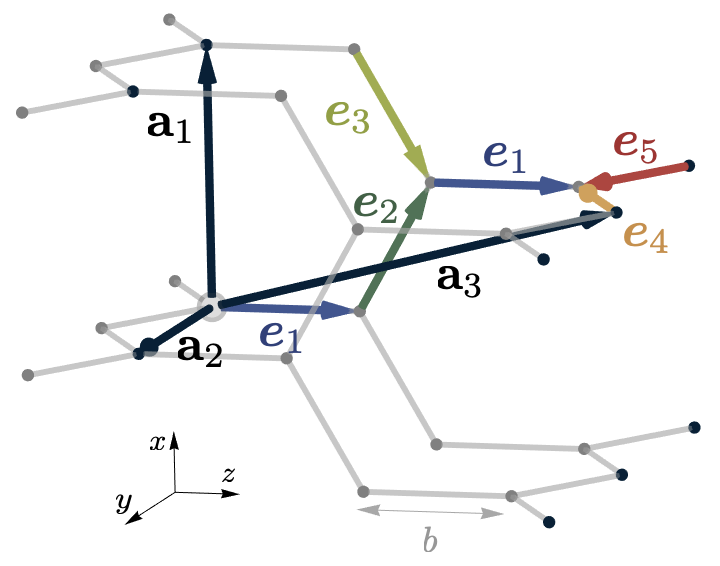} 
    \caption{The hyperhoneycomb lattice defined by the links in Eq.~\eqref{eq:HHClinks}, with the colored links highlighting a unit cell.}
    \label{fig:HHCbaselatt}
\end{figure}

The HHC lattice is defined in terms of five link directions ${\bs e}_\alpha$ and three lattice vectors ${\bf a}_i$,  which are given by in units of the lattice spacing $b$,
\begin{align}
    {\bs e}_1 
    & = (0,0,1) \ , \ {\bs e}_2 = \frac{1}{2}(\sqrt{3},0,1) \ , \ {\bs e}_3 = \frac{1}{2}(-\sqrt{3},0,1) 
    \ , \nn \\
    {\bs e}_4 
    & = \frac{1}{2}(0,-\sqrt{3},-1) \ , \ {\bs e}_5 = \frac{1}{2}(0,\sqrt{3},-1) \ , 
    \label{eq:HHClinks} \\
    {\bf a}_1 & = (\sqrt{3},0,0) \ , \ {\bf a}_2 = (0,\sqrt{3},0) \ , \ {\bf a}_3 = \frac{1}{2}(\sqrt{3},\sqrt{3},6) 
    \ , \nn
\end{align}
with any point on the lattice being generated by shifting the unit cell along the lattice vectors, ${\bs x}=n  {\bf a}_1 + m {\bf a}_2 + l {\bf a}_3$, 
with $n,m,l\in \mathbb{Z}$.
Figure~\ref{fig:HHCbaselatt} shows the 3D lattice emerging from these definitions of links and vertices.
This convention can be related to other definitions found in the literature, e.g., in Refs.~\cite{Mandal:2009,Takayama:2015,OBrien:2016,Jahromi_2021}, by rescalings and rotations.

For the loops we consider, the origin of the coordinate system $(0,0,0)$ is placed at the location of the highlighted vertex in Fig.~\ref{fig:HHCbaselatt}.  
The relevant  geometric quantity in this case is the volume of a single cell, 
\begin{equation}
    V_{\cellHHC}=|({\bf a}_1\times{\bf a}_2)\cdot {\bf a}_3|=9b^3 \ .
\end{equation}

\subsection{Magnetic contribution}
\label{subsec:HHCmag}
\begin{figure}[tb!]
    \centering
    \includegraphics[width=\linewidth]{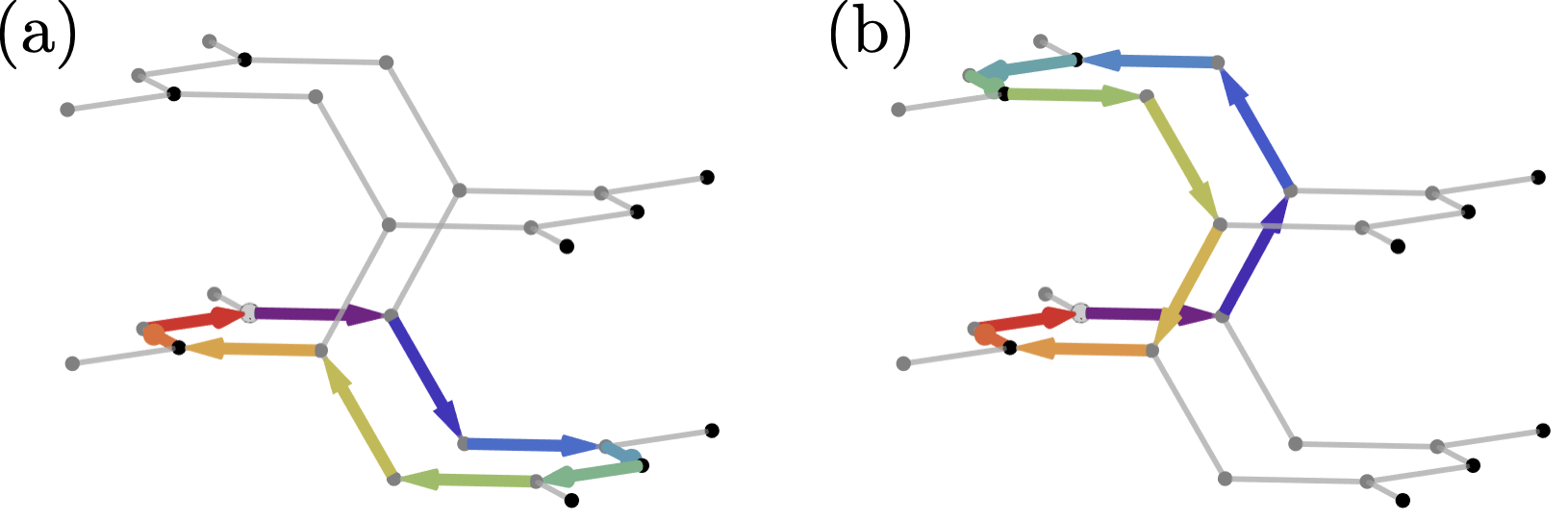} 
    \caption{Examples of (a) 10-link
    and (b) 12-link loops on the HHC lattice.}
    \label{fig:HHC1012}
\end{figure}

\noindent
The shortest loop involves 10 links, shown in Fig.~\ref{fig:HHC1012}(a),
while the next shortest loop involves 12 links,
shown in Fig.~\ref{fig:HHC1012}(b).
Explicit evaluations of these two kinds of loops 
(see Appendix \ref{app:HHC}) 
are required to recover the correct leading-order result. 
That is because both the 10- and 12-link loops are still within a unit cell (which is not the case for the other lattices).
The eight 10-link loops and the four 12-link loops shown in 
Figs.~\ref{fig:HHC10} and \ref{fig:HHC12}, 
and presented in Eqs.~\eqref{eq:HHC10vert} and \eqref{eq:HHC12vert}, respectively,
give
\begin{align}
\Gamma_{10}^{({\rm HHC})} & = \frac{1}{N_{\Phi}} \sum_i\ {\rm Tr} \left[ \hat{I} - e^{-ig\Phi_i} \right] \label{eq:10linkcont}\\
& = \frac{ g^2 b V_{\cellHHC}}{8} {\rm Tr} \left[ G_{xy}^2 + 6 G_{xz}^2 + 6 G_{yz}^2 \right] + {\cal O}(b^6) \ , \nn
\end{align}
where $\Phi_i$ is the contribution from a given loop and $N_\Phi=8$ loops are utilized
(out of the possible 10 loops in a cell),
and
\begin{equation}
\Gamma_{12}^{({\rm HHC})} = \frac{g^2 b V_{\cellHHC}}{2}  {\rm Tr} \left[ G_{xy}^2  \right] + {\cal O}(b^6)
\ .
\label{eq:12linkcont}
\end{equation}
The different numerical coefficients for $G_{ij}$ in Eqs.~\eqref{eq:10linkcont} and~\eqref{eq:12linkcont} are due to the lattice geometry, not the specific gauge-fixing choice.
Combining them together gives
\begin{align}
\Gamma_{10,12}^{({\rm HHC})} & = \Gamma_{10}^{({\rm HHC})} + \frac{5}{4} \Gamma_{12}^{({\rm HHC})} \nn\\
& = \frac{3 g^2 b V_{\cellHHC}}{4} {\rm Tr} \left[ G_{xy}^2 +  G_{xz}^2 +  G_{yz}^2 \right] + {\cal O}(b^6) \nn\\
& = \frac{3 g^2 b V_{\cellHHC}}{8} \sum_{a,i} [B^a_i ({\bs x})]^2 + {\cal O}(b^6) \  .
\label{eq:1012linkcont}
\end{align}
The contributions that do arise at ${\cal O}(b^5)$ in both Eqs.~\eqref{eq:10linkcont} and~\eqref{eq:12linkcont} can all be written in terms of total derivatives, 
 e.g., $D_i \left(G_{jk} G_{lm}\right)$,
and therefore are not shown.
While gauge-invariant, some of these operators are not Lorentz invariant, reflecting the underlying lattice structure.

At ${\cal O}(b^6)$, there are a number of different gauge-invariant operator structures 
present in this combination of loops.
Forming combinations of 14-link and higher loops to cancel these classical contributions 
has eluded us so far.
However, the contribution in Eq.~\eqref{eq:1012linkcont} is improved by one order in the lattice spacing expansion.
At this order, the contribution to the magnetic interaction in the Hamiltonian becomes,
\begin{equation}
H_B  =  \frac{V_{\cellHHC}}{2}\sum_{a,i,{\bs x}}[B^a_i({\bs x})]^2 \ \rightarrow \  \frac{4}{3 g^2 b}\ \sum_{\rm tiles} \Gamma_{10,12}^{({\rm HHC})} 
\ \ .
\label{eq:classHHCMag}
\end{equation}

In 3+1D, it is well known that quantum fluctuations will give rise to further leading-order lattice spacing dependence, as has been well studied in lattice QCD calculations,
requiring tadpole improvement~\cite{Shakespeare:1998uu,Alford:1995hw,Alford:1996pk,Fiebig:1996wg,Poulis:1997zx,Alford:1997mx}, 
as discussed in Sec.~\ref{sec:HC}.
The tadpole improvement from the mean-field value of the plaquette operators can be determined from a single 10-link loop (or computed individually from each loop geometry),
e.g., 
\begin{equation}
u_{\loopHHCb}^{10} = 1 + \frac{1}{2 N_c} \langle\psi | \hat{P}_{10} + \hat{P}^\dag_{10}| \psi \rangle \ .
\label{eq:classHHCtads}
\end{equation}
In terms of the plaquette operators,
the magnetic contribution to the Hamiltonian, given in Eq.~\eqref{eq:classHHCMag}, becomes, 
\begin{align}
H_B =
\frac{2}{3 g^2 b}\ \sum_{\rm tiles} 
& \left[ \frac{9}{2} N_c - \frac{1}{u_{\loopHHCb}^{10}} \left( \hat{P}_{10} + \hat{P}_{10}^\dagger \right) \right. \nn\\
& \quad  \left. - \ \frac{5}{4 u_{\loopHHCb}^{12}} \left( \hat{P}_{12} + \hat{P}_{12}^\dagger \right) \right]
\ ,
\label{eq:classHHCMagP}
\end{align}
where the summation extends over the HHC unit cells, and  $\hat P_{n}$
denotes the average over contributing plaquettes in a given cell, as discussed above.
Corrections to the relation 
in Eq.~(\ref{eq:classHHCMagP})
due to lattice artifacts start at ${\cal O}(b^2)$.

\subsection{Electric contribution}
\label{subsec:HHCe}
\noindent
The six links in a fundamental set of links involved are 
defined in Eq.~\eqref{eq:HHClinks},
and shown in Fig.~\ref{fig:HHCbaselatt} (notice the repetition of $\bs e_1$ to complete the unit cell).
The embedding of the Cartesian electric field,
${\bs E}=(E_x, E_y, E_z)$,
is
\begin{align}
E_1 & = E_z \ , & & \nn\\
E_2 & =  \frac{1}{2} ( \sqrt{3} E_x + E_z) \ ,\ 
& E_3 &= \frac{1}{2} (-\sqrt{3} E_x + E_z) \ , \nn \\
 E_4 & =  -\frac{1}{2} (\sqrt{3} E_y + E_z) \ ,\ 
& E_5 &= \frac{1}{2} (\sqrt{3} E_y - E_z) \ .
\label{eq:EHHC}
\end{align}
The Cartesian electric field is thus given in terms of the link fields by
\begin{align}
|{\bs E}|^2 & = E_x^2+E_y^2+E_z^2 \nn \\
& = 
\frac{2}{3} \sum_{\alpha\neq 1}
\left(E_\alpha\right)^2
 + \frac{1}{6} \sum_{\alpha=1,1'}  \left(E_\alpha\right)^2
\ ,
\label{eq:EHHCcart}
\end{align}
where the difference between $\alpha=1$ and $\alpha=1'$ is to include both $\bs e_1$ links within the unit cell.
Note that this sum involves contributions from different locations in the HHC unit cell.
Therefore, the electric Hamiltonian is given by, after the appropriate KS rescaling $E_\alpha \rightarrow gE_\alpha/b^2$, 
\begin{equation}
H_E = 
\frac{g^2  V_{\cellHHC}}{b^4}\ 
\sum_{{\rm cells}}
\left(
\frac{2}{3}
\sum_{\alpha\neq 1}
+
\frac{1}{6}
\sum_{\alpha=1,1'} 
\right)
{\rm Tr}\!\left[(E_\alpha )^2\right]
\ .
\label{eq:HHCloE}
\end{equation}
This relation has lattice spacing artifacts that start at ${\cal O}(b^2)$, as discussed in the previous section.
Given that the magnetic Hamiltonian has been improved to this order only,  
no further improvement of the electric Hamiltonian is required for the complete 
Hamiltonian,
the sum of 
$H_B$ given in Eq.~(\ref{eq:classHHCMagP}) and $H_E$ given in Eq.~(\ref{eq:HHCloE}),
to be improved with lattice spacing artifacts starting at ${\cal O}(b^2)$.

\section{Summary and Outlook}
\label{sec:Summary}
\noindent
Advances in quantum computing have brought us to the point where 
2+1D and 3+1D
simulations of the dynamics of fundamental quantum systems are becoming possible.
The NISQ era has seen great progress in simulating 1+1D systems, including the Schwinger model,
and the development of algorithms that can scale to arbitrary size simulations for confining and gapped theories~\cite{Farrell:2023fgd,Farrell:2024fit}.   
These techniques are also applicable in 2+1D and 3+1D.

A major challenge for simulations of non-Abelian gauge theories is the complexity of the group algebra, and particularly the rearrangement of group structure in applications of the plaquette operator.  Cubic lattices require recoupling six group spaces at each vertex for the application of plaquette operator due to the structure of the spatial grid.  
Efforts have been made to circumvent this complexity, including the suggestion of working with tessellations of space with only three-link vertices, such as the honeycomb~\cite{Muller:2023nnk} or the triamond~\cite{Kavaki:2024ijd} lattice.  
We have pursued this direction further and developed the improved Kogut-Susskind Hamiltonian for the honeycomb lattice in 2+1D and the hyperhoneycomb lattice in 3+1D.
We were able to include both classical and quantum improvements to both lattices, providing a technique to improve simulations using HC lattice by two orders in the lattice spacing, and those using HHC lattices by one order.

In order to quantify the expected reductions in  resources (either in qubit/qudit count or gate count) for quantum simulations performed using these tri-coordinated lattices versus the cubic lattices, an explicit implementation of these mappings is required for faithful comparisons. 
This is left for future work, together with the inclusion of fermions.

As Euclidean-space lattice field theory simulations of low-lying properties of hadrons and their interactions are significantly more mature than the corresponding quantum simulations, explorations of the HC and HHC lattices could be accelerated through classical lattice simulations.
This includes the development of associated low-energy effective field theories, such as chiral perturbation theory, encoding the underlying symmetries of these tessellations.

\begin{acknowledgments}
\noindent
We would like to thank Randy Lewis for his inspiring talk given to the IQuS group~\cite{randyIQuS}, and S\"oren Schlicting for emphasizing the usage of classical lattice simulations to accelerate the development of these new lattices.
Martin Savage would like to thank 
the High-Energy Physics group
at Universit\"at Bielefeld for kind hospitality during some of this work.
This work was supported, in part, by the Quantum Science Center (QSC)\footnote{\url{https://www.qscience.org}} which is a National Quantum Information Science Research Center of the U.S.\ Department of Energy (Marc), and by U.S.\ Department of Energy, Office of Science, Office of Nuclear Physics, InQubator for Quantum Simulation (IQuS)\footnote{\url{https://iqus.uw.edu/}} under Award Number DOE (NP) Award DE-SC0020970 via the program on Quantum Horizons: QIS Research and Innovation for Nuclear Science\footnote{\url{https://science.osti.gov/np/Research/Quantum-Information-Science}} 
(Martin, Xiaojun).
This work is also supported, in part, through the Department of Physics\footnote{\url{https://phys.washington.edu}} and the College of Arts and Sciences\footnote{\url{https://www.artsci.washington.edu}} at the University of Washington.
We have made extensive use of Wolfram {\tt Mathematica}~\cite{Mathematica}.

\end{acknowledgments}

\bibliography{bib_main}


\appendix

\section{Explicit calculations of the HC loops}
\label{app:HC}
\begin{figure}[tbh!]
    \centering
    \includegraphics[width=\linewidth]{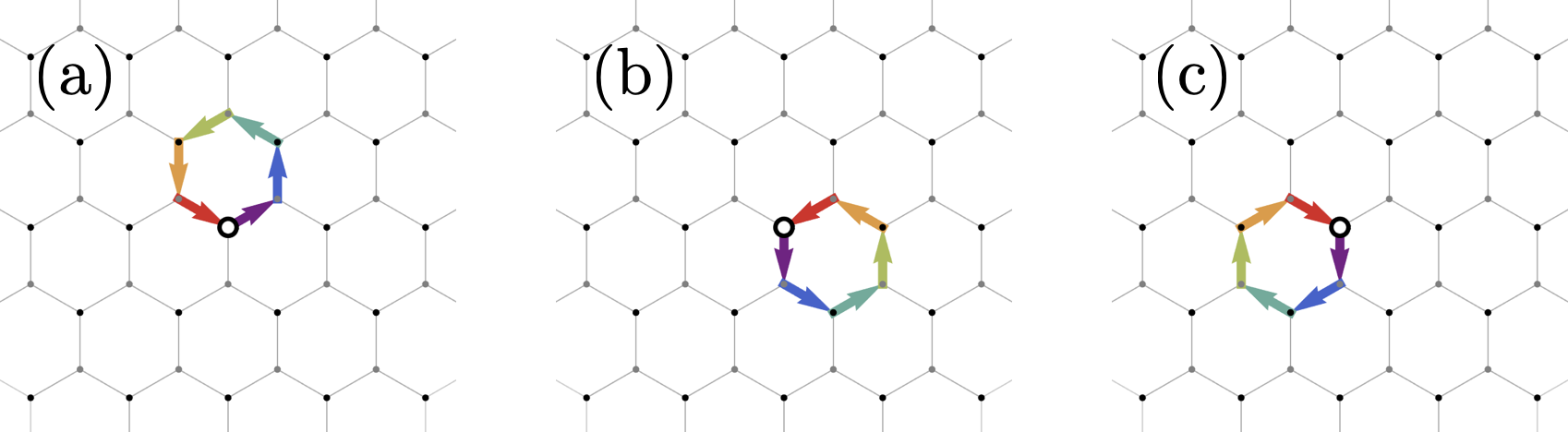} 
    \caption{The 6-link loops on the HC lattice that are averaged together to provide the leading-order contribution to the magnetic Hamiltonian. The (a)-(c) panels show the three different variants about a given lattice site.}
    \label{fig:HC6}
\end{figure}

\noindent
The 6-link loops that are averaged in Eq.~\eqref{eq:P6exp} to furnish the leading contribution to the Yang-Mills Hamiltonian using a honeycomb lattice are shown in Fig.~\ref{fig:HC6}, with their link directions in Eq.~\eqref{eq:HC6vert} 
(starting from the point defined to be $(0,0)$),
\begin{align}
l^{(6)}_a = \{ & {\bs e_1}, -{\bs e_3}, {\bs e_2}, -{\bs e_1}, {\bs e_3}, -{\bs e_2} \} \ , \nn\\
l^{(6)}_b = \{ & {\bs e_3}, -{\bs e_2}, {\bs e_1}, -{\bs e_3}, {\bs e_2}, -{\bs e_1} \} \ , \nn\\
l^{(6)}_c = \{ & {\bs e_3}, -{\bs e_1}, {\bs e_2}, -{\bs e_3}, {\bs e_1}, -{\bs e_2} \} 
\ .
\label{eq:HC6vert}
\end{align}

\begin{figure}[tbh!]
    \centering
    \includegraphics[width=\linewidth]{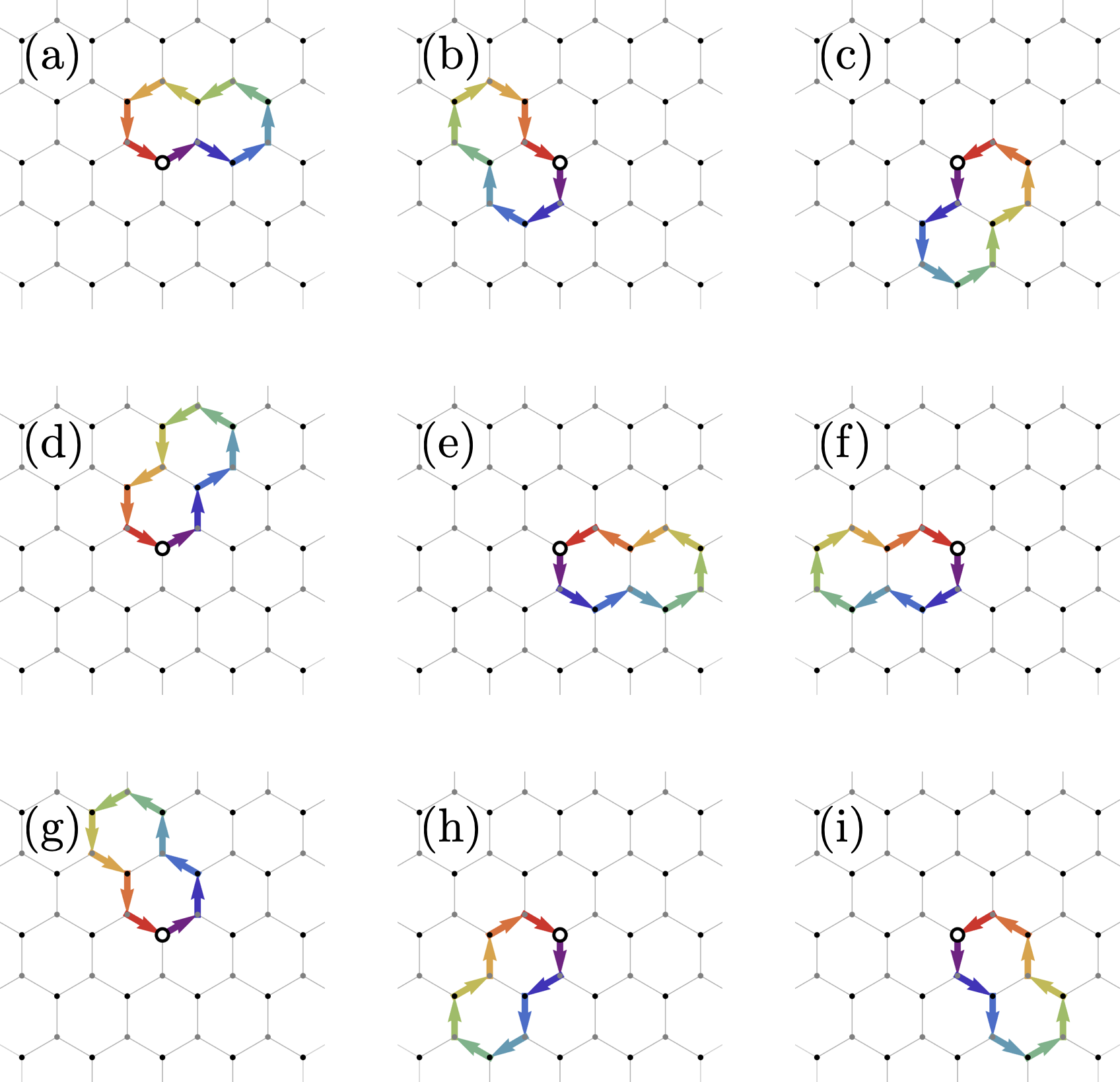} 
    \caption
    {The 10-link loops on the HC lattice used to improve the magnetic contribution to the Hamiltonian. The (a)-(i) panels show the nine different variants about a given lattice site.}
    \label{fig:HC10}
\end{figure}

Classical improvement of the Hamiltonian is accomplished by combining the contributions from the 6-link loops with those from the average of 10-link loops, shown in Fig.~\ref{fig:HC10}, with the corresponding link directions in Eq.~\eqref{eq:HC10vert}, 
\begin{align}
l^{(10)}_{a} = \{ & {\bs e}_1, -{\bs e}_2, {\bs e}_1, -{\bs e}_3, {\bs e}_2, -{\bs e}_1 , \nn \\ &  {\bs e}_2, -{\bs e}_1, {\bs e}_3, -{\bs e}_2 \} \ , \nn\\
l^{(10)}_{b} = \{ & {\bs e}_3, -{\bs e}_1, {\bs e}_2, -{\bs e}_3, {\bs e}_2, -{\bs e}_3 , \nn \\ &  {\bs e}_1, -{\bs e}_2, {\bs e}_3, -{\bs e}_2 \} \ , \nn\\
l^{(10)}_{c} = \{ & {\bs e}_3, -{\bs e}_1, {\bs e}_3, -{\bs e}_2, {\bs e}_1, -{\bs e}_3 , \nn \\ &  {\bs e}_1, -{\bs e}_3, {\bs e}_2, -{\bs e}_1 \} \ , \nn\\
l^{(10)}_{d} = \{ & {\bs e}_1, -{\bs e}_3, {\bs e}_1, -{\bs e}_3, {\bs e}_2, -{\bs e}_1 , \nn \\ &  {\bs e}_3, -{\bs e}_1, {\bs e}_3, -{\bs e}_2 \} \ , \nn\\
l^{(10)}_{e} = \{ & {\bs e}_3, -{\bs e}_2, {\bs e}_1, -{\bs e}_2, {\bs e}_1, -{\bs e}_3 , \nn \\ &  {\bs e}_2, -{\bs e}_1, {\bs e}_2, -{\bs e}_1 \} \ , \nn\\
l^{(10)}_{f} = \{ & {\bs e}_3, -{\bs e}_1, {\bs e}_2, -{\bs e}_1, {\bs e}_2, -{\bs e}_3 , \nn \\ &  {\bs e}_1, -{\bs e}_2, {\bs e}_1, -{\bs e}_2 \} \ , \nn\\
l^{(10)}_{g} = \{ & {\bs e}_1, -{\bs e}_3, {\bs e}_2, -{\bs e}_3, {\bs e}_2, -{\bs e}_1 , \nn \\ &  {\bs e}_3, -{\bs e}_2, {\bs e}_3, -{\bs e}_2 \} \ , \nn\\
l^{(10)}_{h} = \{ & {\bs e}_3, -{\bs e}_1, {\bs e}_3, -{\bs e}_1, {\bs e}_2, -{\bs e}_3 , \nn \\ &  {\bs e}_1, -{\bs e}_3, {\bs e}_1, -{\bs e}_2 \} \ , \nn\\
l^{(10)}_{i} = \{ & {\bs e}_3, -{\bs e}_2, {\bs e}_3, -{\bs e}_2, {\bs e}_1, -{\bs e}_3 , \nn \\ &  {\bs e}_2, -{\bs e}_3, {\bs e}_2, -{\bs e}_1 \} 
\ .
\label{eq:HC10vert}
\end{align}

The operator coefficients in Eq.~\eqref{eq:HCloop} derived from the 6-link and 10-link loops are displayed in Table~\ref{tab:HCcoeffs}.
\begin{table}[H]
\renewcommand{\arraystretch}{1.4}
    \centering
    \begin{tabularx}{\linewidth}{||Y|c|c|c|c|c|c||}
    \hline
         Loop & $\alpha$ & $\beta_1$ & $\beta_2$ & $\gamma_1$ & $\gamma_2$ & $\gamma_3$
         \\ \hline
         6-link & & & & & & \\ \hline
         Fig.~\ref{fig:HC6}(a)  & $1$  & $0$             & $1$    & $5/48$    & $0$             & $29/48$  \\
         Fig.~\ref{fig:HC6}(b)  & $1$  & $\sqrt{3}/2$    & $-1/2$ & $23/48$   & $-\sqrt{3}/4$   & $11/48$ \\
         Fig.~\ref{fig:HC6}(c)  & $-1$ & $\sqrt{3}/2$    & $1/2$  & $-23/48$  & $-\sqrt{3}/4$   & $-11/48$ \\ \hline
         10-link & & & & & & 
         \\ \hline
         Fig.~\ref{fig:HC10}(a) &  $2$  & $\sqrt{3}$    & $2$   & $41/24$  & $\sqrt{3}$     & $29/24$ \\
         Fig.~\ref{fig:HC10}(b) &  $-2$ & $3\sqrt{3}/2$ & $-1/2$& $-25/12$ & $3 \sqrt{3}/4$ & $-5/6$ \\
         Fig.~\ref{fig:HC10}(c) &  $2$  & $\sqrt{3}/2$  & $-5/2$& $7/12$   & $-\sqrt{3}/4$  & $7/3$ \\
         Fig.~\ref{fig:HC10}(d) &  $2$  & $\sqrt{3}/2$  & $7/2$ & $7/12$   & $5 \sqrt{3}/4$ & $23/6$ \\
         Fig.~\ref{fig:HC10}(e) &  $2$  & $2\sqrt{3}$   & $-1$  & $95/24$  & $-\sqrt{3}$    & $11/24$ \\
         Fig.~\ref{fig:HC10}(f) &  $-2$ & $2\sqrt{3}$   & $1$   & $-95/24$ & $-\sqrt{3}$    & $-11/24$ \\
         Fig.~\ref{fig:HC10}(g) &  $2$  & $-\sqrt{3}/2$ & $7/2$ & $7/12$   & $-5 \sqrt{3}/4$& $23/6$ \\
         Fig.~\ref{fig:HC10}(h) &  $-2$ & $3\sqrt{3}/2$ & $5/2$ & $-25/12$ & $-9 \sqrt{3}/4$& $-7/3$ \\
         Fig.~\ref{fig:HC10}(i) &  $2 $ & $3\sqrt{3}/2$ & $-5/2$ & $25/12$ & $-9 \sqrt{3}/4$& $7/3$ \\ \hline
    \end{tabularx}
    \caption{Numerical values of the operator coefficients in Eq.~\eqref{eq:HCloop} for the 6- and 10-link loops in the honeycomb lattice.}
    \label{tab:HCcoeffs}
\end{table}
%

\section{Explicit calculations of the HHC loops}
\label{app:HHC}
\noindent
\begin{figure}[tbh!]
    \centering
    \includegraphics[width=\linewidth]{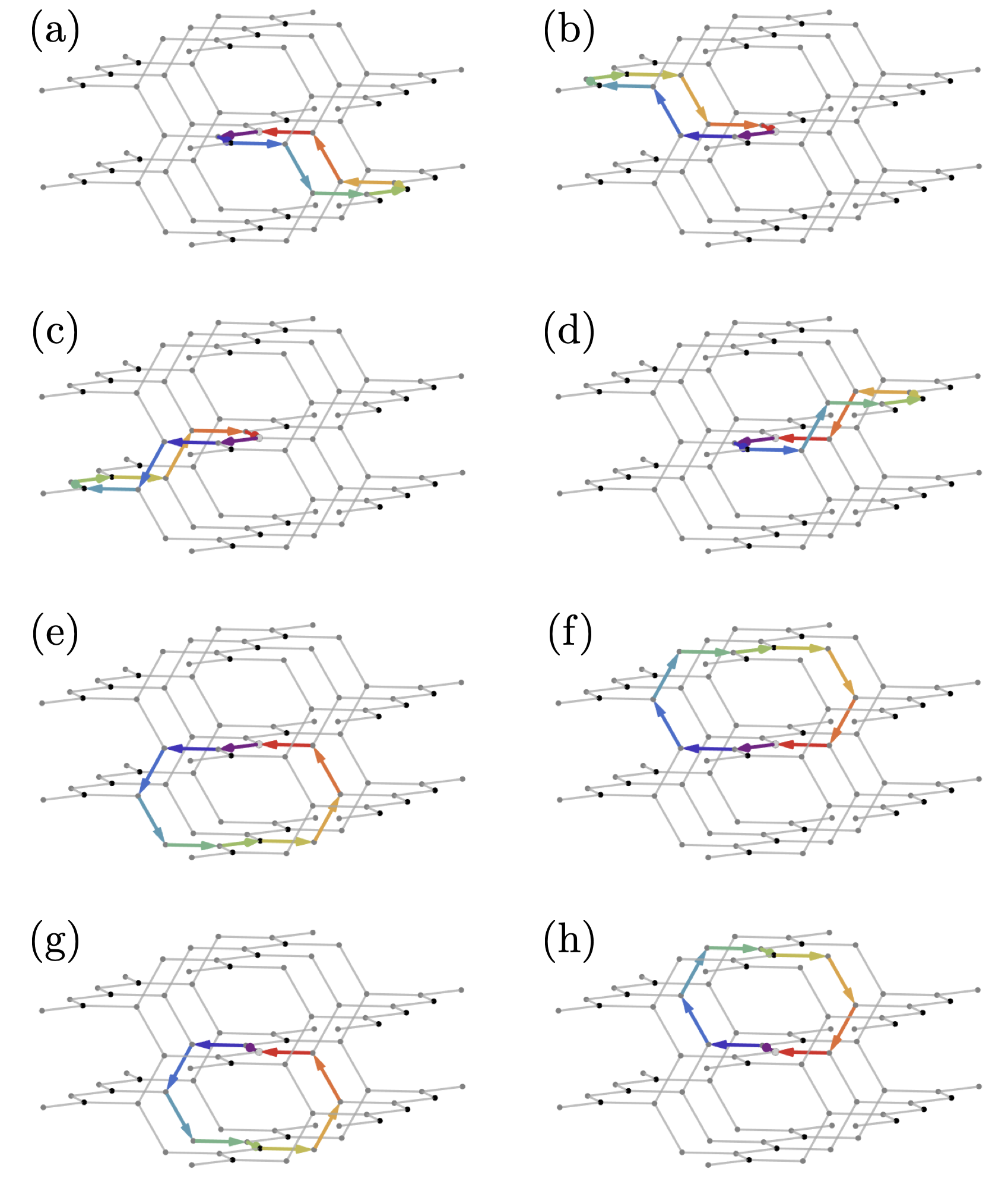} 
    \caption{The 10-link loops of the HHC lattice contributing to the magnetic Hamiltonian. The (a)-(h) panels show the eight different variants about a given lattice site.}
    \label{fig:HHC10}
\end{figure}
\begin{figure}[tbh!]
    \centering
    \includegraphics[width=\linewidth]{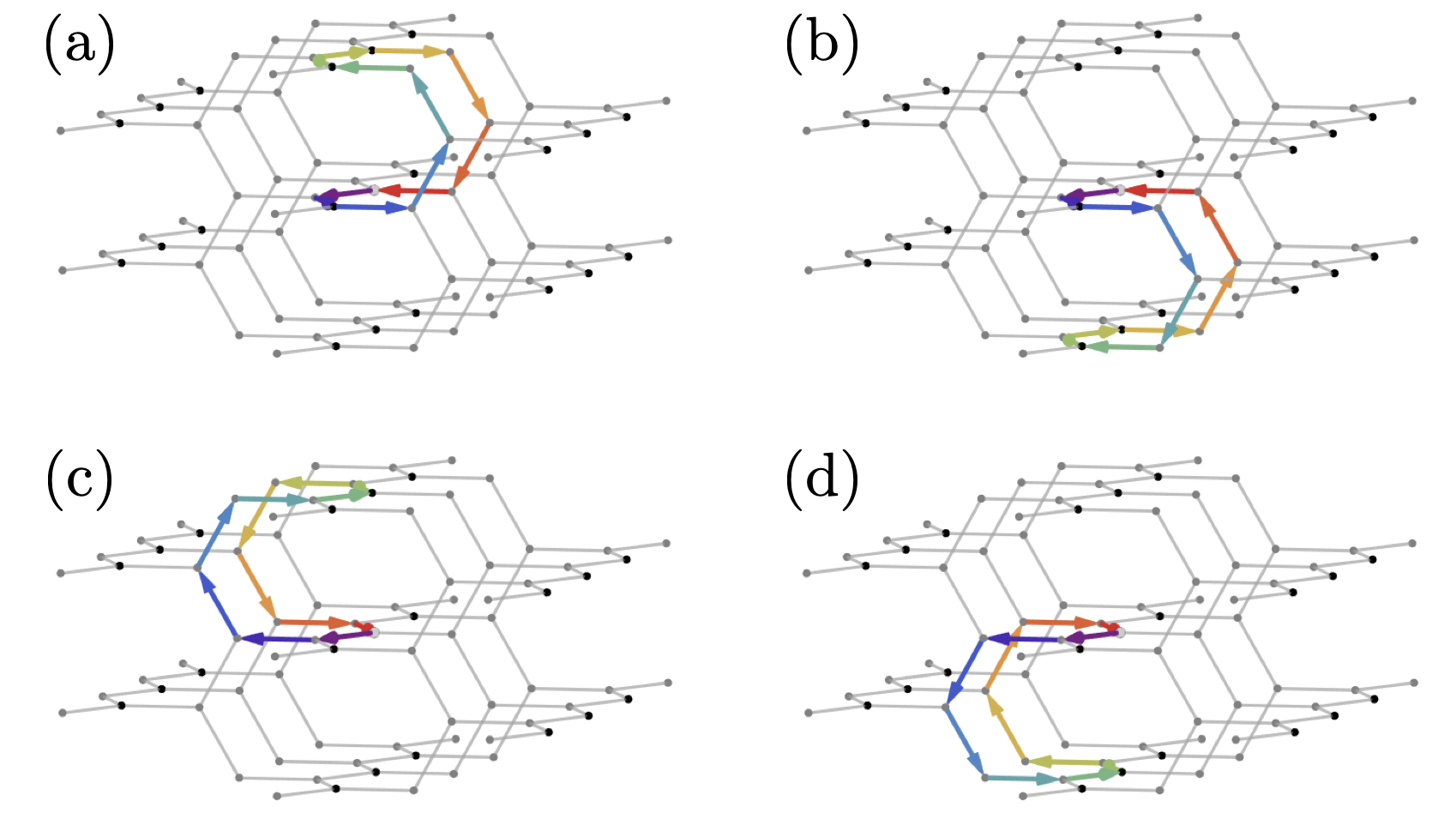} 
    \caption{The 12-link loops of the HHC lattice contributing to the magnetic Hamiltonian. The (a)-(d) panels show the four different variants about a given lattice site.}
    \label{fig:HHC12}
\end{figure}

The eight 10-link loops used to obtain Eq.~\eqref{eq:10linkcont} are shown in Fig.~\ref{fig:HHC10}, with the corresponding link directions in Eq.~\eqref{eq:HHC10vert}, 
\begin{align}
l^{(10)}_{a} = \{ & {\bs e}_5, -{\bs e}_4, {\bs e}_1, {\bs e}_3, {\bs e}_1, -{\bs e}_5,   \nn \\ & {\bs e}_4, -{\bs e}_1, -{\bs e}_3, -{\bs e}_1 \}\ , \nn \\
l^{(10)}_{b} = \{ & {\bs e}_5, -{\bs e}_1, -{\bs e}_3, -{\bs e}_1, {\bs e}_4, -{\bs e}_5, \nn \\ &  {\bs e}_1, {\bs e}_3, {\bs e}_1, -{\bs e}_4 \}\ , \nn \\
l^{(10)}_{c} = \{ & {\bs e}_5, -{\bs e}_1, -{\bs e}_2, -{\bs e}_1, {\bs e}_4, -{\bs e}_5, \nn \\ &  {\bs e}_1, {\bs e}_2, {\bs e}_1, -{\bs e}_4 \}\ , \nn \\
l^{(10)}_{d} = \{ & {\bs e}_5, -{\bs e}_4, {\bs e}_1, {\bs e}_2, {\bs e}_1, -{\bs e}_5,   \nn \\ & {\bs e}_4, -{\bs e}_1, -{\bs e}_2, -{\bs e}_1 \}\ , \nn \\
l^{(10)}_{e} = \{ & {\bs e}_5, -{\bs e}_1, -{\bs e}_2, {\bs e}_3, {\bs e}_1, -{\bs e}_5,  \nn \\ & {\bs e}_1, {\bs e}_2, -{\bs e}_3, -{\bs e}_1 \}\ , \nn \\
l^{(10)}_{f} = \{ & {\bs e}_5, -{\bs e}_1, -{\bs e}_3, {\bs e}_2, {\bs e}_1, -{\bs e}_5,  \nn \\ & {\bs e}_1, {\bs e}_3, -{\bs e}_2, -{\bs e}_1 \}\ , \nn \\
l^{(10)}_{g} = \{ & {\bs e}_4, -{\bs e}_1, -{\bs e}_2, {\bs e}_3, {\bs e}_1, -{\bs e}_4,  \nn \\ & {\bs e}_1, {\bs e}_2, -{\bs e}_3, -{\bs e}_1 \}\ , \nn \\
l^{(10)}_{h} = \{ & {\bs e}_4, -{\bs e}_1, -{\bs e}_3, {\bs e}_2, {\bs e}_1, -{\bs e}_4,  \nn \\ & {\bs e}_1, {\bs e}_3, -{\bs e}_2, -{\bs e}_1 \}\ ,
\label{eq:HHC10vert}
\end{align}
while
the 12-link loops used to obtain Eq.~\eqref{eq:12linkcont} are shown in Fig.~\ref{fig:HHC12}, with the corresponding link directions in Eq.~\eqref{eq:HHC12vert}, 
\begin{align}
l^{(12)}_{a} = \{ & {\bs e}_5, -{\bs e}_4, {\bs e}_1, {\bs e}_2, -{\bs e}_3, -{\bs e}_1, \nn \\
& {\bs e}_4, -{\bs e}_5, {\bs e}_1, {\bs e}_3, -{\bs e}_2, -{\bs e}_1 \}\ , \nn \\
l^{(12)}_{b} = \{ & {\bs e}_5, -{\bs e}_4, {\bs e}_1, {\bs e}_3, -{\bs e}_2, -{\bs e}_1, \nn \\
& {\bs e}_4, -{\bs e}_5, {\bs e}_1, {\bs e}_2, -{\bs e}_3, -{\bs e}_1 \}\ , \nn \\
l^{(12)}_{c} = \{ & {\bs e}_5, -{\bs e}_1, -{\bs e}_3, {\bs e}_2, {\bs e}_1, -{\bs e}_5, \nn \\
& {\bs e}_4, -{\bs e}_1, -{\bs e}_2, {\bs e}_3, {\bs e}_1, -{\bs e}_4 \}\ , \nn \\
l^{(12)}_{d} = \{ & {\bs e}_5, -{\bs e}_1, -{\bs e}_2, {\bs e}_3, {\bs e}_1, -{\bs e}_5, \nn \\
& {\bs e}_4, -{\bs e}_1, -{\bs e}_3, {\bs e}_2, {\bs e}_1, -{\bs e}_4 \}
\ .
\label{eq:HHC12vert}
\end{align}

To help the reader visualize these 3-dimensional loops, we provide a \texttt{Mathematica} file as Supplemental Material with the 10-link and 12-link loops plotted as 3D plots, with the possibility to rotate and view them in different angles.

As in the case of the HC loops, the HHC loops can also be written as $P(\bs x) = {\rm Tr}[e^{ig\Phi(\bs x)}]$, with $\Phi(\bs x)$ having the following general form,
\begin{align}
    \Phi(\bs x) = &\ b^2\left[\alpha_1 G_{xy}(\bs x)+\alpha_2 G_{zx}(\bs x)+\alpha_3 G_{zy}(\bs x) \right. \nn \\
    & +b(\beta_1 D_x + \beta_2 D_y + \beta_3 D_z)G_{xy}(\bs x)  \nn \\
    & +b(\beta_4 D_x + \beta_5 D_y + \beta_6 D_z)G_{zx}(\bs x)  \nn \\
    & \left. + \, b (\beta_7 D_y + \beta_8 D_z)G_{zy}(\bs x) +\mathcal{O}(b^2) \right] \ ,
    \label{eq:HHCloop}
\end{align}
where the relation from Eq.~\eqref{eq:gaugegzy} has been used to remove the term $D_xG_{zy}$.
The numerical values of $\alpha_i$ and $\beta_i$ are shown in Tables~\ref{tab:HHCcoeffsAlfa} and \ref{tab:HHCcoeffsBeta}, respectively.
\begin{table}[H]
\renewcommand{\arraystretch}{1.4}
    \centering
    \begin{tabularx}{\linewidth}{||Y|Y|Y|Y||}
    \hline
         Loop & $\alpha_1$ & $\alpha_2$ & $\alpha_3$
         \\ \hline
         10-link & & &
         \\ \hline
         Fig.~\ref{fig:HHC10}(a) &  $3/2$  & $0$          & $-3\sqrt{3}$    \\
         Fig.~\ref{fig:HHC10}(b) &  $-3/2$ & $0$          & $3\sqrt{3}$ \\
         Fig.~\ref{fig:HHC10}(c) &  $3/2$  & $0$          & $3\sqrt{3}$ \\
         Fig.~\ref{fig:HHC10}(d) &  $-3/2$ & $0$          & $-3\sqrt{3}$  \\
         Fig.~\ref{fig:HHC10}(e) &  $3/2$  & $3\sqrt{3}$  & $0$   \\
         Fig.~\ref{fig:HHC10}(f) &  $-3/2$ & $-3\sqrt{3}$ & $0$    \\
         Fig.~\ref{fig:HHC10}(g) &  $-3/2$ & $3\sqrt{3}$  & $0$  \\
         Fig.~\ref{fig:HHC10}(h) &  $3/2$  & $-3\sqrt{3}$ & $0$  \\ \hline
         12-link & & & 
         \\ \hline
         Fig.~\ref{fig:HHC12}(a) &  $-3$ & $0$ & $0$    \\
         Fig.~\ref{fig:HHC12}(b) &  $3$  & $0$ & $0$ \\
         Fig.~\ref{fig:HHC12}(c) &  $-3$ & $0$ & $0$ \\
         Fig.~\ref{fig:HHC12}(d) &  $3$  & $0$ & $0$  \\ \hline
    \end{tabularx}
    \caption{Numerical values of the operator coefficients $\alpha_i$ in Eq.~\eqref{eq:HHCloop} for the 10- and 12-link loops in the hyperhoneycomb lattice.}
    \label{tab:HHCcoeffsAlfa}
\end{table}
\begin{table}[H]
\renewcommand{\arraystretch}{1.4}
    \centering
    \begin{tabularx}{\linewidth}{||Y|c|c|c|c|c|c|c|c||}
    \hline
         Loop & $\beta_1$ & $\beta_2$ & $\beta_3$ & $\beta_4$ & $\beta_5$ & $\beta_6$ & $\beta_7$ & $\beta_8$ 
         \\ \hline
         10-link & & & & & & & &
         \\ \hline
         Fig.~\ref{fig:HHC10}(a) &  $-\frac{3 \sqrt{3}}{8}$ & $\frac{3 \sqrt{3}}{4}$ & $-\frac{33}{8}$& 0 & $-\frac{15}{8}$ & 0 & $-\frac{9}{2}$ & $-\frac{15 \sqrt{3}}{4}$ \\
         Fig.~\ref{fig:HHC10}(b) &  $-\frac{3 \sqrt{3}}{8}$ & 0                    & $-\frac{39}{8}$& 0 & $-\frac{21}{8}$ & 0 & 0 & $-\frac{21 \sqrt{3}}{4}$ \\
         Fig.~\ref{fig:HHC10}(c) &  $-\frac{3 \sqrt{3}}{8}$ & 0                    & $\frac{39}{8}$ & 0 & $\frac{21}{8}$ & 0 & 0 & $-\frac{21 \sqrt{3}}{4}$ \\
         Fig.~\ref{fig:HHC10}(d) &  $-\frac{3 \sqrt{3}}{8}$ & $-\frac{3 \sqrt{3}}{4}$& $\frac{33}{8}$ & 0 & $\frac{15}{8}$ & 0 & $-\frac{9}{2}$ & $-\frac{15 \sqrt{3}}{4}$ \\
         Fig.~\ref{fig:HHC10}(e) &  $-\frac{3 \sqrt{3}}{4}$ & $\frac{3 \sqrt{3}}{8}$ & $\frac{3}{8}$  & $-\frac{9}{2}$ & $\frac{21}{8}$ & $-\frac{3 \sqrt{3}}{4}$ & 0 & 0 \\
         Fig.~\ref{fig:HHC10}(f) &  $-\frac{3 \sqrt{3}}{4}$ & $-\frac{3 \sqrt{3}}{8}$& $-\frac{3}{8}$ & $-\frac{9}{2}$ & $-\frac{21}{8}$ & $\frac{3 \sqrt{3}}{4}$ & 0 & 0 \\
         Fig.~\ref{fig:HHC10}(g) &  $\frac{3 \sqrt{3}}{4}$  & $\frac{3 \sqrt{3}}{8}$ & $-\frac{3}{8}$ & $-\frac{9}{2}$ & $-\frac{21}{8}$ & $-\frac{3 \sqrt{3}}{4}$ & 0 & 0 \\
         Fig.~\ref{fig:HHC10}(h) &  $\frac{3 \sqrt{3}}{4}$  & $-\frac{3 \sqrt{3}}{8}$& $\frac{3}{8}$  & $-\frac{9}{2}$ & $\frac{21}{8}$ & $\frac{3 \sqrt{3}}{4}$ & 0 & 0 \\ \hline
         12-link & & & & & & & &
         \\ \hline
         Fig.~\ref{fig:HHC12}(a) & $-\tfrac{3\sqrt{3}}{2}$ & $-\tfrac{3\sqrt{3}}{2}$ & $-\tfrac{3}{4}$ & 0 & $\tfrac{15}{4}$ & 0 & 0 & 0 \\
         Fig.~\ref{fig:HHC12}(b) & $-\tfrac{3\sqrt{3}}{2}$ & $-\tfrac{3\sqrt{3}}{2}$ & $\tfrac{3}{4}$  & 0 & $-\tfrac{15}{4}$& 0 & 0 & 0 \\
         Fig.~\ref{fig:HHC12}(c) & $-\tfrac{3\sqrt{3}}{2}$ & 0                     & $-\tfrac{3}{4}$ & 0 & $-\tfrac{21}{4}$  & 0 & 0 & 0 \\
         Fig.~\ref{fig:HHC12}(d) & $-\tfrac{3\sqrt{3}}{2}$ & 0                     & $\tfrac{3}{4}$  & 0 & $\tfrac{21}{4}$   & 0 & 0 & 0 \\ \hline
    \end{tabularx}
    \caption{Numerical values of the operator coefficients $\beta_i$ in Eq.~\eqref{eq:HHCloop} for the 10- and 12-link loops in the hyperhoneycomb lattice.}
    \label{tab:HHCcoeffsBeta}
\end{table}


\end{document}